\documentclass[fleqn,twoside]{article}
\usepackage{espcrc2}

\usepackage{graphicx}
\usepackage[figuresright]{rotating}

\title{Science with the Square Kilometer Array: 
Motivation, Key Science Projects, Standards and Assumptions}

\author{C.L. Carilli\address{National Radio Astronomy Observatory,
    Socorro, NM, 87801, USA, ccarilli@nrao.edu},
 S. Rawlings\address{Astrophysics, Department of Physics, Keble Road, 
 Oxford, OX1 3RH, UK, s.rawlings1@physics.ox.ac.uk}}
       
\begin{document}

\begin{abstract}

The Square Kilometer Array (SKA) represents the next major, and
natural, step in radio astronomical facilities, providing two orders of
magnitude increase in collecting area over existing telescopes.  In a
series of meetings, starting in Groningen, the Netherlands (August
2002) and culminating in a `science retreat' in Leiden (November
2003), the SKA International Science Advisory Committee (ISAC),
conceived of, and carried-out, a complete revision of the SKA science
case (to appear in New Astronomy Reviews). This preface includes: (i)
general introductory material, (ii) summaries of the key science
programs, and (iii) a detailed listing of standards and assumptions
used in the revised science case.

\end{abstract}

\maketitle

\section{Process and Intent}

The 1970's and 1980's saw the construction of numerous fully steerable
single dish and array radio telescopes with apertures of
order $10^4 ~ \rm m^2$. These telescopes allowed the study of HI 21cm
emission from gas-rich galaxies in the nearby universe, e.g.\ out
to the Virgo cluster, and, with effort, to redshift $z \sim 0.2$. In the
subsequent 25 years, while optical astronomy has seen an order-of-magnitude 
increase in collecting area for ground-based telescopes,
collecting area in radio astronomy has remained stagnant.

The idea for a `Square Kilometer Array' (SKA) was born in the early
1990's.  Such a telescope would provide two orders of magnitude
increase in collecting area\footnotemark  over existing telescopes, 
allowing for
study of the HI content of galaxies to cosmologically significant
distances (i.e.\ to $z \sim 2$ rather than $z \sim 0.2$). 
The SKA project was born as an international program with
the establishment by the International Union of Radio Science (URSI)
in September 1993 of the Large Telescope Working Group.  

\footnotetext{ 
For radio telescopes (unlike optical telescopes) this leads to a
two-dex increase in line, e.g.\ HI, sensitivity. For continuum observations,
the SKA will be three-dex more sensitive than current synthesis arrays, largely
because of increased bandwidth; the EVLA and e-MERLIN projects are designed
to `bridge the gap' by increasing the bandwidth of the existing VLA
and MERLIN arrays.
}

The first science case for the SKA was edited by R. Taylor and
R. Braun, and appeared as a publication from the Netherlands
Foundation for Radio Astronomy in 1999.  In August 2000, at the
International Astronomical Union meeting in Manchester (UK), a
Memorandum of Understanding to establish the International Square
Kilometre Array Steering Committee (ISSC) was signed by
representatives of eleven countries (Australia, Canada, China,
Germany, India, Italy, the Netherlands, Poland, Sweden, the United
Kingdom, and the United States). More information, and the current
timeline, for the SKA project can be found at
http://www.skatelescope.org.

The time since the publication of the Taylor-Braun document has seen a
revolution in our knowledge of the local and distant Universe. 
We have entered an era of `precision cosmology', where the fundamental
parameters ($H_{0}$, $\Omega_{\rm M}$ etc)
describing the emerging `standard model' in cosmology are known to 
$\sim \pm 10 \%$. This standard model includes `dark energy'
and `dark matter' as the two dominant energy densities in the present-day
Universe. We have probed into the time of first light in universe, the
`epoch of reionization', when the UV emission from the first stars and
(accreting) supermassive black holes reionizes the neutral
intergalactic medium.  Gamma-ray bursts have been shown to be the
largest explosions in the universe, tracing the death of very massive
stars to the earliest epochs.  Supermassive black holes have gone from
being a hypothetical bi-product of general relativity (GR), to being a
fundamental aspect of all spheroidal galaxies and how these
objects formed. Galactic `micro-quasars', or accreting black holes of a few
solar masses, have been shown to have similar properties to their
supermassive cousins, but on eight orders of magnitude smaller mass scales.
The recent discovery of the first known double pulsar presents the 
promise of the most accurate tests of strong-field GR. 
Extra-solar planets are now known to be a common phenomenon, and the search 
for terrestrial planets has begun. A new constituent of the solar system 
has been confirmed, the Kuiper belt objects, and these may provide the key to
understanding the formation of the solar nebula.  Wide-field
surveys, and ultra-deep narrow-field surveys, from radio through X-ray
wavelengths, have mapped out our Galaxy and the large-scale structure of the
local Universe in astonishing detail.

In a series of meetings, starting in Groningen, the Netherlands 
(August 2002) and culminating in a `science retreat' in Leiden
(November 2003), the SKA International Science Advisory Committee
(ISAC), conceived of, and carried-out, a complete revision of the SKA
science case. This incorporated the newest results in astronomy, with
emphasis on the most important outstanding problems. 
The revised science case is organized along the lines
of nine science working groups (Table 1), covering all areas of modern
astrophysics. The chairs of these working groups enlisted authors
from a broad spectrum of the community, including theorists and
multi-wavelength observers. The aim of the working groups was for
completeness, with discussion of all areas in which the SKA will play
a pivotal r\^{o}le in advancing our knowledge.  However,
the authors were instructed not to simply present review articles in
radio astronomy. The resulting articles emphasize detailed analyses of topical
research programs where the unique capabilities of the SKA can be
exploited. These analyses employ the latest simulations and
theoretical models of physical phenomena, such as cosmic reionization,
galaxy formation, and star and planet formation. As such, these
articles represent original research, and the book should act as an
important reference for the planning of SKA `path-finder' projects like
as LOFAR and HYFAR, as well as enhancements to existing facilities like
the EVLA and e-MERLIN.

\begin{table*}[htb]
\small
\caption{The ISAC Working Groups} 
\vskip 0.2in
\begin{tabular}{cc}
\hline
\hline
Area  & Chair \\
\hline
The Milky Way and Local Galaxies & John Dickey (Minnesota) \\
SETI, Stellar End Products, Transient Sources & Joseph Lazio (NRL) \\
Cosmology and Large Scale Structure & Frank Briggs (ANU) \\
Galaxy Evolution & Thijs van der Hulst (Kapteyn) \\
Active Galactic Nuclei and Super Massive Black Holes &  Heino Falcke 
(ASTRON) \\
The Life Cycle of Stars & Sean Dougherty (DRAO) \\
The Solar System and Planetary Science & Bryan Butler (NRAO) \\
The Intergalactic Medium & Luigina Feretti (IRA) \\
Spacecraft Tracking & Dayton Jones (JPL) \\
\hline
\end{tabular}
\end{table*}

\section{Key Science Projects}

In parallel with work on the science book, where the emphasis is on
completeness, the ISAC has recognized a few `Key Science Projects' (KSPs) for
the SKA.  These projects are defined according to three criteria: (i)
ability to address important but currently unanswered questions in
fundamental physics or astrophysics, (ii) science which is either
unique to the radio band and the SKA, or is complementary to other
wavebands, but in which the SKA plays a key r\^{o}le, and (iii) excites
the broader community, and is of relevance and interest to funding
agencies.

A sub-committee of the ISAC, chaired by Bryan Gaensler, was created to
establish the KSPs for the SKA.  Proposals were solicited
from the community, and debated within the ISAC at the Geraldton SKA
(August 2003) and at the Leiden science retreat (November
2003). A consensus was reached by the ISAC to put-forward five KSPs
for the SKA to the International SKA Steering
Committee (ISSC). These were approved by the ISSC, and appeared in
February 2004 as SKA memo 44. The first five articles in this volume 
present each of these KSPs in detail.

\begin{itemize}

\item {\bf KSP I. The Cradle of Life} There is increasing interest in 
astrobiology and the search for Earth-like planets.  The
SKA has the unique potential for studying extra-solar
terrestrial planet formation, and for detecting signals from
other life like us. At 20~GHz, the SKA will
provide thermal imaging at 0.15-AU resolution out to a distance of
150~pc, encompassing many of the best studied Galactic star-forming
regions.  Such observations will allow us to study the process of {\em
terrestrial} planet formation, as well as studies of the evolution
these proto-planetary disks on sub-AU scales on timescales of months
(``movies of planet formation''). The SKA will have the capability of
detecting `leakage radiation' from extraterrestial intelligence (ETI)
transmitters associated with the nearest stars, and targeted searches will
involve studies of up to a million solar-type stars. Finally, the
SKA will have the resolution and sensitivity to study the low order
transitions of amino acids and other complex carbon biomolecules, and
to follow their progress from molecular clouds to protoplanets.

\item {\bf KSP II. Strong-Field Tests of Gravity Using Pulsars and Black Holes:}
Pulsar surveys with the SKA can discover tens of thousands of pulsars,
amongst which there is an excellent chance of finding
a pulsar in orbit around a black hole which will yield the first measurements 
of relativistic gravity in the ultra-strong-field limit.
Thousands of millisecond pulsars will also be discovered 
which can form an immense `pulsar timing array' with which gravitational
waves may be detected. New probes of GR opened up will include tests of the
Cosmic Censorship Conjecture and the No-Hair theorem.

\item {\bf KSP III. The Origin and Evolution of Cosmic Magnetism:} 
Radio astronomy is uniquely placed in its capability to study magnetic
fields at large distances, through studies of Faraday rotation,
polarized synchrotron emission and the Zeeman effect. 
The SKA could measure rotation measures for $\sim10^8$ polarized 
extragalactic sources across an entire hemisphere, with an average spacing
between sightlines of $\sim 60$ arcsec. The sheer weight of statistics in
these data, combined with deep spectropolarimetric observations of nearby
galaxies and clusters will allow a complete characterization of the
evolution of magnetic fields in galaxies and clusters from redshifts
$z>3$ to the present. It will be possible to determine whether there is a 
connection between the formation of magnetic fields and the formation of
structure in the early Universe, and to provide solid constraints on
when and how the first magnetic fields in the Universe were generated.

\item {\bf KSP IV. Galaxy Evolution and Cosmology:} The original
motivation for building the SKA was to detect HI in normal galaxies at
high redshift, and recent developments in galaxy evolution and
cosmology make this an extremely exciting prospect The SKA will
provide the only means of studying the cosmic evolution of neutral
Hydrogen (HI) which, alongside information on star formation from the
radio continuum, is needed to understand how stars formed from gas
within dark-matter over-densities. `All hemisphere' HI redshift
surveys to $z \sim 1.5$ are feasible with wide-field-of-view
realisations of the SKA, and, by measuring the galaxy power spectrum
in exquisite detail, will allow the first precise studies of the
equation-of-state of dark energy.  The SKA will be capable of other
uniquely powerful cosmological studies including the measurement of
the dark-matter power spectrum using weak gravitational lensing, and
the precise measurement of $H_{0}$ using extragalactic water masers.

\item {\bf KSP V. Probing the Dark Ages:} The epoch of reionization (EoR),
during which the first luminous objects in the universe formed, and
then reionized the neutral IGM, can only be studied at near-IR through
radio wavelengths.  The SKA provides fundamental probes of the EoR in
two critical areas. First, the ability of the SKA to image the neutral
IGM in HI 21cm emission (and absorption) is a  unique probe of
the process of reionization, and is recognized as the next fundamental
step in our study of the evolution of large scale structure and cosmic
reionization. Second, the incomparable sensitivity of the SKA
enables studies of the molecular gas, dust, and star formation
activity in the first galaxies, as well as the radio continuum
emission from the first accreting supermassive black holes.

\end{itemize}

An important aspect of defining KSPs is their impact on SKA
array specification and design.  Table 2 summarizes the SKA design
specifications dictated by the the current\footnotemark  KSPs. Each of the
main SKA specifications can be traced-back to one or more of the 
KSP requirements. This table is discussed at length in SKA memo 44.

\footnotetext{
Both the KSPs and their SKA design implications need to be continuously
scrutinised to ensure that the SKA is positioned to deliver the best possible science;
an example of a possible modification to the implied specifications for the
Galaxy Evolution and Cosmology KSP is given as a footnote to Table 2.
}

\begin{table*}[htb]
\scriptsize
\caption{SKA specifications implied by Key Science Project goals.}
\vskip 0.2in
\begin{tabular}{lcccc}
\hline
\hline
Topic & $A_{\rm eff}/ T_{\rm sys}$ &  Frequencies & Max Baseline  & Special \\   
 & (m$^2$~K$^{-1}$) & (GHz) & (km) &   \\
\hline
Gravity & 20\,000 at 1.4 GHz & 0.5--15  & 3000 & 
multifielding desirable? (TBD);  \\
~      & timing array & Galactic Center & astrometry & significant 
central core \\ \hline
Dark Ages & 10\,000 at 0.1 MHz \& 20 GHz & 0.1--20 & 3000 & 35~GHz for CO
studies; \\
~ & CO emission at $z>6$ (M~82) & ~ & HI absorption & central core for HI; \\
~ & HI structure at $6< z<13$ & ~ & SMBH studies & full FOV imaging at 1.4
GHz \\  \hline
Magnetism & 20\,000 at 1.4 GHz & 0.3--10 & 300 &  --40~dB polarizationn purity; \\
~       &  RM grid & large RMs & confusion-limited  & central core; \\
   & & & imaging at 1.4 GHz & full FOV imaging at 1.4 GHz \\
\hline
Cradle of Life & 10\,000 at 20 GHz & $\ge 20$ & 3000 & 100 pencil beams
within \\
~       & 10~K rms at 1~mas & terrestrial planet & 0.15 AU at 150 pc &
FOV for targeted searches; \\
~       & in 100 hrs & formation &  at 20 GHz & central core \\
\hline
Evolution   & 20\,000 at 1.4 GHz & 0.3--1.4\footnotemark & 300 &  dedicated beam with
FOV of \\
\& LSS       & M$_*$ galaxy at $z=2$ & galaxies to $z=4$ & ~ &
~200 deg$^2$ 
at 0.7~GHz is highly \\
 & & & & desirable to increase survey speed \\
\hline
\hline
\end{tabular}
\end{table*}

\footnotetext{
It is plausible that a precise measurement of $H_{0}$ via SKA observations of extragalactic
water masers might prove a key experiment in early-21$^{\rm st}$-century cosmology, requiring 
an SKA that works up to $\sim 22 ~ \rm GHz$ with high line-sensitivity on intercontinental baselines.
}

There was a long debate within the ISAC concerning whether or not `Exploration of the Unknown'
constituted a sixth KSP. This volume ends with a detailed discussion of this topic.

\section{Standards and Assumptions}

A critical aspect of a work such as this is to establish strict
telescope parameters which all authors adopt in their scientific
calculations. In August 2003 the SKA project office appointed
Dayton Jones to codify and rationalize the SKA specifications, in
consultation with the ISAC and the Engineering Management Team (EMT).
The results are shown below (see SKA memo 45).  All the
science articles adopt these specifications as nominal. The authors
were also asked to point out areas where small changes to these specifications
could lead to substantial improvements in the science delivered.

\begin{itemize}

\item	Frequency range:     100 MHz - 25 GHz     Goal: 60 MHz - 35 GHz

\item 	Simultaneous independent observing bands:		
		2 pairs  (2 polarizations at each of two independent 
		frequencies, with same FoV centers)

\item	Maximum frequency separation of observing bands: 
		Factor of 3 between observing
		band center frequencies (same FoV centers)

\item 	Instantaneous bandwidth of each observing band:	Full width = 25\% 
		of observing band center frequency, up to a maximum of 4 
		GHz BW for all frequencies above 16 GHz

\item 	Sensitivity at 45 degrees elevation ($A_{\rm eff} / T_{\rm sys}$ in units of  m$^2$/K):
		(Goal:  2500 at 60 MHz)
		5000 at 200 MHz, 
		20000 between 0.5 and 5 GHz, 
		15000 at 15 GHz, and
		10000 at 25 GHz
		(Goal:  5000 at 35 GHz) 
(see Table 3 for dual polarization sensitivities).

\item 	Configuration:	Minimum baselines 20 meters,  20\% of total
		collecting area within 1 km diameter,  50\% of total 
		collecting area within 5 km diameter,  75\% of total 
		collecting area within 150 km diameter,  maximum 
		baselines at least 3000 km from array core
		(angular resolution $< 0.02$ / f GHz arcsec)

\item 	Image quality:	Dynamic range $> 10^6$ and image fidelity $> 10^4$
		between 0.5 and 25 GHz, over a range of 90 degrees in 
		declination and 100 in angular resolution

\item 	Contiguous imaging field of view (FoV):	  1 square degree 
		within half power points at 1.4 GHz, scaling as 
		wavelength squared,  goal: 200 square degree field of view 
		within half power points at 0.7 GHz, scaling as
		wavelength squared between 0.5-1.0 GHz

\item 	Number of separated fields of view:	1 with full sensitivity.
					Goal:	4 with full sensitivity.
					10 simultaneous sub-arrays

\item	Correlator and post-correlation processing:
		Input bandwidth 25\% of center frequency for 
			frequencies below 16 GHz and 4 GHz for 
			frequencies above 16 GHz (per observing band).
		Imaging of 1 square degree at 1.4 GHz with 0.1 arcsec 
			angular resolution.
		Imaging of 200 sq. degrees at 0.7 GHz with 0.2 arcsec 
			angular resolution.
		Imaging of $10^4$ separate regions within the FoV, each 
			covering at least $10^5$ beam areas at full 
			(maximum baseline) angular resolution.
		Spectral resolution of $10^4$ channels per observing 
			band per baseline.
		Minimum sampling interval 0.1 ms for wide-field pulsar 
			searches

\item	Beamformer capability:	50 simultaneous summed (phased array) 
		beams within FoV, inner 5 km diameter of  array.  
		No time averaging, 8 bits/sample.

\item	Survey speed\footnotemark:	
		FoV x (A/T)$^2$ x BW  =  $3\times 10^{17}$ deg$^{2}$ m$^4$ K$^{-2}$ Hz$^{-1}$ at 1.5 GHz;
		FoV x (A/T)$^2$ x BW  =  $1.5 \times 10^{19}$ deg$^{2}$ m$^4$ K$^{-2}$ Hz$^{-1}$ at 0.7 GHz

\item	Antenna pointing and slewing:  	Blind pointing $< 0.1$ HPBW, 
		move between adjacent sky positions separated by 
		0.5 HPBW in 3 sec, move between sky positions sep. 
		by 90 deg. in $< 60$ s

\item	Instrumental polarization:	Polarization error / total 
		intensity -40 dB at FoV center, -30 dB out to FoV edge 
		(after routine calibration)

\item	Spectral dynamic range:  10$^4$  (flatness of bandpass response 
		after calibration)

\item	Total power calibration:	Total power (zero-spacing) 
		flux density measured with 5\% error within 1 hr.

\end{itemize}

\footnotetext{
Survey speed here means the speed with which a fixed sky area is surveyed to a
fixed limiting flux density.
}

\begin{table}[htb]
\caption{SKA rms sensitivity in 1 hour}
\begin{tabular}{ccc}
\hline
\hline
Frequency & Bandwidth & rms \\
GHz & MHz & $\mu$Jy \\
\hline
0.2 & 50 & 1.4 \\
1.4 & 350 & 0.13 \\
8 & 2000 & 0.06 \\
20 & 4000 & 0.08 \\
\hline
\hline
\end{tabular}
\end{table}

For calculations concerning cosmologically distant sources, a standard
`concordance cosmology' was adopted with $H_{0} \simeq 70 ~ \rm km \,
s^{-1} \, Mpc^{-1}$, $\Omega_M \simeq 0.3$, $\Omega_\Lambda \simeq
0.7$, and $\Omega_B \simeq 0.04$ (and $\sigma_8 \simeq 1$ and $n_{\rm
scalar} \simeq 1$ for the normalisation and shape of the matter power
spectrum) unless stated otherwise.

We were unable (or perhaps unwilling) to enforce uniformity over
important trivia like systems of units, sign conventions (e.g.\ on
radio spectral index $\alpha$) and American versus the Queen's
English.

\section{Acknowledgements}

The editors would like to thank the members of the ISAC, and in
particular, the chairs of the working groups, for their efforts in
writing the SKA science case.  This was truly a team effort, and would
never have been completed without substantial organizational and
scientific input from the working groups. 

Likewise, we thank the authors for their contributions. It is clear
from reading the articles that the authors took seriously the charge
to present substantive, original, and relevant research programs that
can be accomplished with the SKA. We encourage all to continue these
exciting lines of research.

We thank the ISSC, and in particular the two ISSC chairs,
Ron Ekers and Jill Tarter, for their support (and protection!) during
the time this work was undertaken. And lastly, we thank the project
director, Richard Schilizzi, for guidance and leadership throughout
the process. 

We acknowledge support from the NRAO, the US SKA consortium, the
Lorentz Centre in Leiden, the SKA Project Office, Oxford University
and the Australia Telescope National Facility. SR is grateful to the
UK PPARC for a Senior Research Fellowship.

\end{document}